\documentclass[a4paper,fleqn,usenatbib]{mnras}
\usepackage{newtxtext,newtxmath}

\usepackage[T1]{fontenc}
\usepackage{ae,aecompl}

\usepackage{amsmath}	
\usepackage{amssymb}	
\usepackage{graphicx}	

\usepackage{natbib}
\usepackage{longtable}
\usepackage{color}
\newcommand{\asec}{$^{\prime\prime}$}

\def\r1415{$^{14}$N/$^{15}$N}

\def\H{N$_{2}$H$^{+}$}
\def\D{N$_{2}$D$^{+}$}
\def\15N{$^{15}$NNH$^+$}
\def\N15{N$^{15}$NH$^+$}

\def\kms{\mbox{km~s$^{-1}$}}
\def\cmc{cm$^{-3}$}
\def\cmq{cm$^{-2}$}

\def\Tex{\mbox{$T_{\rm ex}$}}

\title[D and $^{15}$N fractionation in \H ]{Deuterium and $^{15}$N fractionation in \H\ during the formation of a Sun-like star
\thanks{Based on observations carried out with the IRAM 30m Telescope.
IRAM is supported by INSU/CNRS (France), MPG (Germany) and IGN (Spain).}}

\author[De Simone et al.]{M. De Simone,$^{1}$\thanks{E-mail: marta.de.smn@gmail.com}
            F. Fontani,$^{2}$
            C. Codella,$^{2}$
           C. Ceccarelli,$^{3}$ 
            B. Lefloch,$^{3}$
            R. Bachiller,$^{4}$ \newauthor
            A. L\'opez-Sepulcre,$^{3,5}$
            E. Caux,$^{6,7}$
            C. Vastel,$^{6,7}$
            J. Soldateschi$^{1}$
          \\
$^{1}$Dipartimento di Fisica e Astronomia, Universit\`a degli Studi di Firenze, I-50125 Firenze, Italy \\
$^{2}$INAF-Osservatorio Astrofisico di Arcetri, Largo E. Fermi 5, I-50125, Florence, Italy \\
$^{3}$Univ. Grenoble Alpes, CNRS, IPAG, F-38000 Grenoble, France \\
$^{4}$Observatorio Astron\'omico Nacional (OAN, IGN), Calle Alfonso XII, 3, 28014 Madrid, Spain \\
$^{5}$Institut de Radioastronomie Millim\'etrique, 300 rue de la Piscine, 38406, Saint-Martin d'H\`eres, France \\
$^{6}$Universit\'e de Toulouse, UPS-OMP, IRAP, Toulouse, France \\
$^{7}$CNRS, IRAP, 9 Av. Colonel Roche, BP 44346, F-31028 Toulouse Cedex 4, France \\
          }

\date{Accepted XXX. Received YYY; in original form ZZZ}

\pubyear{2018}

\begin{document}
\label{firstpage}
\pagerange{\pageref{firstpage}--\pageref{lastpage}}
\maketitle

\begin{abstract}
Although chemical models predict that the deuterium fractionation in \H\ is a good 
evolutionary tracer in the star formation process, the fractionation of nitrogen 
is still a poorly understood process. Recent models have questioned the similar evolutionary 
trend expected for the two fractionation mechanisms in \H , based on a
classical scenario in which ion-neutral reactions occurring in cold gas should have
caused an enhancement of the abundance of \D , \15N , and \N15. In the framework of the ASAI IRAM-30m large program, we have investigated 
the fractionation of deuterium and $^{15}$N in \H\ in the best known representatives of the 
different evolutionary stages of the Sun-like star formation process. The goal is to ultimately
confirm (or deny) the classical "ion-neutral reactions" scenario that predicts a similar trend 
for D and $^{15}$N fractionation. We do not find any evolutionary trend of the \r1415\ ratio
from both the \15N\ and \N15\ isotopologues. Therefore, our findings confirm that,
during the formation of a Sun-like star, the core evolution is irrelevant in the fractionation of 
$^{15}$N. The independence of the \r1415\ ratio with time, found also in high-mass 
star-forming cores, indicates that the enrichment in $^{15}$N revealed in comets and 
protoplanetary disks is unlikely to happen at core scales. Nevertheless, we have firmly 
confirmed the evolutionary trend expected
for the H/D ratio, with the \H /\D\ ratio decreasing before the pre--stellar core phase, and
increasing monotonically during the protostellar phase. We have also confirmed
clearly that the two fractionation mechanisms are not related. 
\end{abstract}


%
\section{Introduction}
\label{intro}

Observations and chemical models agree that the process of deuterium enrichment in 
\H\ is a robust evolutionary indicator in the star formation process. The root ion-neutral
reaction that forms \D\ is (e.g.~Millar et al.~\citeyear{millar}, Ceccarelli et al.~\citeyear{ceccarelli2014}): 
\begin{equation}
{\rm H_3^+ + HD \leftrightarrow H_2D^+ + H_2 + \Delta E} \;, 
\end{equation}
followed by the reaction of ${\rm H_2D^+}$ with ${\rm N_2}$ to give \D . Reaction (1) is 
exothermic by ${\rm \Delta E \sim 230}$~K, and hence it is fast only from left to right below $\sim 20$~K.
At these low temperatures, and if CO is highly depleted, the abundance of ${\rm H_2D^+}$
is thus boosted, and so are the species directly formed from it, like \D . This implies
that the \D /\H\ ratio is expected to decrease from the pre--stellar to the proto-stellar phase,
when the nascent protostar begins to heat up its surrounding material 
(e.g.~Caselli~\citeyear{caselli2002}). In fact, observations of pre-stellar cores and
young protostars have shown D/H ratios in \H\ of the order of $0.1 - 0.01$
(e.g.~Bacmann et al.~\citeyear{bacmann}, Crapsi et al.~\citeyear{crapsi}, Emprechtinger et al.~\citeyear{emprechtinger}), 
i.e. several orders of magnitude higher than the cosmic D/H elemental abundance 
($\sim 10^{-5}$, e.g.~Linsky et al.~\citeyear{linsky}), and a decrease with core 
evolution in both low- and high-mass star-forming cores (Emprechtinger et al.~\citeyear{emprechtinger}, 
Ceccarelli et al.~\citeyear{ceccarelli2014}, Fontani et al.~\citeyear{fontani2011}, Bianchi
et al.~\citeyear{bianchi}). This trend is followed by other molecules formed by ${\rm H_2D^+}$, 
such as DNC (e.g.~Fontani et al.~\citeyear{fontani2014}, Gerner
 et al.~\citeyear{gerner}) and ${\rm DCO^+}$ (e.g.~Emprechtinger et al.~\citeyear{emprechtinger},
Gerner et al.~\citeyear{gerner}). Despite the gradual decrease with time expected after 
the protostar birth, the huge D enrichment in the early cold phase is able to explain 
the enhanced D/H ratio measured in comets ($\sim 10^{-4}$, e.g.~Hartogh et 
al.~\citeyear{hartogh}, Altwegg et al.~\citeyear{altwegg}), with respect to the cosmic 
value (Cleeves et al.~\citeyear{cleeves}). For this reason, the D/H ratio is believed to
be an excellent chemical tool to link the different phases of the Solar system formation,
from the earliest pre--stellar stage to the formation of the planets and other Solar system
bodies (Ceccarelli et al.~\citeyear{ceccarelli2014}).

Similarly to D, $^{15}$N is enriched in pristine Solar system bodies such as comets 
(\r1415 $\sim 150$, Manfroid et al.~\citeyear{manfroid}, Shinnaka et al.~\citeyear{shinnaka}) 
and carbonaceous condrites ($\sim 50$, Bonal et al.~\citeyear{bonal}) by factors $\sim 2$ to 
$\sim 10$ with respect to the protosolar nebula value ($\sim 441$, Marty et 
al.~\citeyear{marty}). As for \D , the \15N\ and \N15\ abundances were thought to be enhanced
through ion-neutral reactions occuring in cold gas (Terzieva \& Herbst~\citeyear{teh}): 
\begin{equation}
{\rm ^{15}N + ^{14}N_2H^+ \leftrightarrow\ ^{14}N + ^{15}NNH^+ + 36 K\;,}
\end{equation}
\begin{equation}
{\rm ^{15}N + ^{14}N_2H^+ \leftrightarrow\ ^{14}N + N^{15}NH^+ + 28 K\;.}
\end{equation}
However, recent chemical models (Roueff et al.~\citeyear{roueff}) have questioned the efficiencies
of these reactions, because of the presence of energy barriers higher than previously predicted, 
which make significant fractionation unlikely to occur even in the earliest cold phases of the 
protostellar evolution. The few observational findings obtained to date make the puzzle even 
more intriguing: the \r1415\ measurements in \H\ in high-mass star-forming cores obtained by
Fontani et al.~(\citeyear{fontani2015}) do not show an evolutionary dependence, in agreement 
with Roueff et al.~(\citeyear{roueff}) predictions, but the same observations indicate a huge dispersion 
of the values (\r1415 $\sim 100 - 1000$) which cannot be reproduced by the models. 
The models also do not predict the high \r1415\ ratio found in the low-mass pre--stellar core 
L1544 ($\sim 1000$, Bizzocchi et al.~\citeyear{bizzocchi}).

In this paper, we show the first evolutionary study of the combined \r1415\ and D/H ratio in \H\ 
in the best representatives of the main evolutionary stages of the Sun-like star formation process.
The primary goal of the study is to understand whether the two ratios are linked and, consequently, 
whether (also) the $^{14}$N/$^{15}$N is an evolution-dependent parameter and can be considered a 
chemical link between the earliest and the most evolved stages during the formation of Sun-like stars. 

\section{Sample and observations}
\label{obs}

The targets are extracted from the Large Program 
ASAI\footnote{http://www.iram-institute.org/EN/content-page-344-7-158-240-344-0.html} 
(Lefloch et al.~in prep.) and represent the best known precursors of a Sun-like star, in the main phases
of its formation: the pre--stellar core L1544, the Class 0 protostars IRAS 4A, 
the Class 0/Class I protostar L1527 and the Class I protostar SVS13-A. We have analysed
also the Class 0 source IRAS 16293 using the data of the TIMASSS survey
(Caux et al.~\citeyear{caux}). Together with this "evolutionary" 
sample, we have studied other sources to provide additional relevant information about the chemical 
evolution in a Sun-like star-forming environment: the protocluster OMC--2 FIR4, the chemically rich 
protostellar shock L1157--B1, and the intermediate-mass Class 0 protostar CepE. 
In the following, we give a brief description of each target (for more details, please see
L\'opez-Sepulcre et al.~\citeyear{lopsep2015}):

\begin{itemize}
\item \textbf{L1544} is a starless core in the Taurus molecular cloud complex 
(d $\sim$140 pc, Cernicharo \& Gu\'elin \citeyear{cernicharo&guelin}). 
It is considered the prototypical pre--stellar core on the verge of the gravitational collapse
(Caselli et al. \citeyear{caselli2012}, and references therein). L1544 is characterized by 
a nucleus with a high H$_2$ density peak ($2\times 10^6$ cm$^{-3}$) and low temperature ($\sim$ 7 K),
surrounded by a lower density envelope undergoing extended infall (Caselli et al.~\citeyear{caselli2012}).
For these reasons, in the central region, CO is depleted by a factor $\sim 10$ 
(Caselli et al.~\citeyear{caselli1999}) and the deuterium fractionation is high, 
like in the interiors of dark clouds, although differentiated chemical processes can take place 
in the external layers (Caselli et al. \citeyear{caselli1999}, Vastel et al. \citeyear{vastel2014}).

\item \textbf{IRAS 16293} is a well known Class 0 protostar located in the small L1689N molecular cloud
in the $\rho$ Ophiuchus complex at a distance of 120 pc (Loinard et al. \citeyear{loinard2008}). 
It is a system with two main components, IRAS16293 A and IRAS16293 B, separated by $\sim$5\asec ,
characterized by a strong chemical differentiation
(Bisschop et al. \citeyear{bisschop2008}; J{\o}rgensen et al. \citeyear{joergensen2011}, \citeyear{joergensen2012}). 
It is the first source where a hot corino has been discovered, 
(Ceccarelli et al. \citeyear{ceccarelli2000b}; Cazaux et al. \citeyear{cazaux}; Bottinelli et al. \citeyear{bottinelli2004}), 
and a well-studied astrochemical laboratory thanks to its richness in complex organic molecules,
and its high deuterium fractionation for example in 
formaldehyde (Ceccarelli et al. \citeyear{ceccarelli1998}, \citeyear{ceccarelli2001}), methanol 
(Parise et al. \citeyear{parise2004}), methyl formate (Demyk et al. \citeyear{demyk2010}), and
water (Coutens et al. \citeyear{coutens2012}).

\item \textbf{IRAS 4A} is the second hot corino ever discovered (Bottinelli et 
al. \citeyear{bottinelli2007}), and is a binary source in the Perseus molecular complex 
(d $\sim$ 235 pc, Hirota et al. \citeyear{hirota2008}, Hirota et al.~\citeyear{hirota2011}); 
it is composed of two Class 0 objects separated by 1$\farcs$8: IRAS4 A1 
and IRAS4 A2 (e.g. Looney et al. \citeyear{looney2000}). 
The nature of IRAS4 A1 and IRAS4 A2 has been discussed in several papers
(e.g.~Persson et al.~\citeyear{persson}, Taquet et al. \citeyear{taquet2015}, Santangelo et al. \citeyear{santangelo2015}, De 
Simone et al. \citeyear{desimone}, L\'{o}pez-Sepulcre et al. \citeyear{lopsep2017}).
In particular, Santangelo et al. (\citeyear{santangelo2015}) have concluded that A1 is
brigther than A2 in the millimeter continuum, but only A2 is associated with a hot-corino.

\item \textbf{L1527} is a dark cloud in the Taurus molecular complex (d $\sim$140 pc), with a heavily
obscured IRAS source (IRAS 04368+2557) located at the core center, classified as a 
borderline Class 0/Class I object according to Andr\'{e} et al. (\citeyear{andre2000}). 
This source is considered as a prototypical warm-carbon-chain-chemistry source 
(WCCC, Sakai et al. \citeyear{sakai2008}, \citeyear{sakai2010}), and high-angular resolution
observations indicate the presence of a compact, infalling, and rotating envelope
(Sakai et al.~\citeyear{sakai2016}, \citeyear{sakai2017}) surrounding the protostar.

\item \textbf{SVS13-A} is a part of the multiple system NGC1333-SVS13
(distance 235~pc, Hirota et al. \citeyear{hirota2008}) in the Perseus complex,
dominated by three millimetre sources identified by interferometric observations 
(Bachiller et al. \citeyear{bachiller1998}, Looney et al. \citeyear{looney2000}, Lefavre
et al.~\citeyear{lefevre}, Tobin et al. \citeyear{tobin2016}, and references therein): 
SVS13-A, SVS13-B separated by $\sim$15\asec and associated with two different evolutionary 
stages, and SVS13-C, $\sim$20\asec away from SVS13-A. Although SVS13-A is still deeply embedded
in a large scale envelope (Lefloch et al. \citeyear{lefloch1998}), it is considered a young Class I 
protostar because of its extended outflow ($>0.07$ pc) and its low ratio between submillimeter
and bolometric luminosity ($\sim$0.8\%), 
becoming an almost unique laboratory to investigate how deuteration and fractionation change from 
the Class 0 to the Class I phases (Chen et al. \citeyear{chen2009}, Tobin et al. \citeyear{tobin2016}, 
Bianchi et.~al~\citeyear{bianchi}, Lefloch et al in prep). In the analysis of SVS13-A, care needs to 
be taken in the fit of the lines because of the contamination of core IRAS 4A at velocity 
$\sim$ 7 \kms\ in the off position (Santangelo et al.~\citeyear{santangelo2015}), 
while that of SVS13-A is at about 8--9 \kms\, 
which results in an artificial absorption near the emission lines of interest. 
This contamination has affected only the border of the blue side of the lines 
(see the figures in the on-line Appendix-A); in the case of the \H\ (1--0) line 
we used the observation carried out in Position Switching mode, and not in Wobbler 
Switching mode, to avoid this contamination.

\item \textbf{OMC--2 FIR4} is a protocluster of low- and intermediate-mass protostars 
at a distance of 420 pc (e.g.~Hirota et al.\citeyear{hirota2007}). Its $\sim$30 M$_{\sun}$ envelope 
extends to about $10^4$ AU (Crimier et al. \citeyear{crimier2009}) and contains several clumps, 
probably a cluster of protostars (Shimajiri et al. \citeyear{shimajiri2008}, L\'opez-Sepulcre et 
al.~\citeyear{lopsep2013}), which makes it one 
of the best and closest environment analogues of the one in which the Sun was born 
(e.g.~Ceccarelli et al.~\citeyear{ceccarelli2014}, Fontani et al.~\citeyear{fontani2017}). In fact,
growing evidence indicates that our Sun was born in a crowded environment, forming together
with many other protostars, including very likely more massive objects (e.g.~Adams~\citeyear{adams},
Taquet et al.~\citeyear{taquet2016}). The fractionation of nitrogen has been studied by
Kahane et al.~(submitted to ApJ) in several molecular species (HCN, HNC, HC$_3$N, CN),
who derived \r1415 $\sim 290$, regardless of the molecule used, and in remarkable agreement 
with the present-day local interstellar \r1415 .

\item \textbf{L1157--B1} is a chemically rich bipolar outflow (Bachiller et al. \citeyear{bachiller1997}, 
\citeyear{bachiller2001}) driven by a low-mass ($\sim$ 4 L$_{\sun}$) Class 0 protostar 
(L1157-mm, at a distance of 250 pc, Looney et al. \citeyear{looney2007}), and associated 
with several blue and red-shifted shocks at different ages (Gueth et al.~\citeyear{gueth1996},~\citeyear{gueth1998}). 
It may be regarded as the ideal laboratory for observing the effects of shocks on the gas chemistry 
(Bachiller et al.~\citeyear{bachiller2001}, Lefloch et al.~\citeyear{lefloch2010},
Codella et al. \citeyear{codella2010}) and on isotopic fractionation (Fontani et al.~\citeyear{fontani2014},
Busquet et al.~\citeyear{busquet2017}). The analysis of the \H\ emission performed by 
Codella et al. (\citeyear{codella2013}) showed that the \H\ emission detected towards the 
L1157-B1 shock can be considered a fossil record of the pre-shock phase, when the gas density was 
$\sim$ 10$^4$ cm$^{-3}$. In other words, \H\ is sampling the chemical-physical 
conditions of the quiescient gas evolved for more than $10^4$ yr, before the gas
was perturbed by the protostellar shock induced by the L1157 protostellar activity.

\item \textbf{CepE} is an intermediate-mass Class 0 protostar (L = 100 L$_{\sun}$, 
Lefloch et al. \citeyear{lefloch1996}) in the Cepheus OB3 association at a distance of 730 pc 
(Sargent \citeyear{sargent1977}), which drives an exceptionally powerful and luminous 
molecular outflow. Both the protostar and the outflow have been intensively 
studied at mm and IR wavelengths (Lefloch et al. \citeyear{lefloch1996}, Moro-Martin et 
al. \citeyear{moro-martin2001}, Noriega-Crespo et al. \citeyear{noriega-crespo2004}).
\end{itemize}

The observations have been performed during several runs between 2011 and 2016
with the IRAM 30-m telescope near Pico Veleta (Spain) in the context of the Astrochemical
Surveys at IRAM (ASAI) Large Program, using the broad-band EMIR receivers E090, 
E150 and E230. These were carried out in Wobbler Switching Mode, with a throw of 3$^\prime$, 
in order to ensure a flat baseline across the spectral bandwidth observed. 
The instrumental setup was decided according to the sources:
L1544 was observed in the band 72 to 115 GHz using the Fast Fourier Transform Spectrometer (FTS)
in its 50 kHz resolution mode; about Protostars and Outflow Shocks, the 3mm (72 -- 116 GHz) and 
2mm (126 -- 170 GHz) bands were observed at the same time. The 1.3mm (200 -- 272 GHz) band was 
covered observing LSB and USB simultaneously, using the FTS in its 200 kHz resolution mode 
(Lefloch et al. in prep.).
The data were reduced with the CLASS software of the GILDAS\footnote{http://www.iram.fr/IRAMFR/GILDAS} 
package. Calibration uncertainties are $\sim$ 10\% at 3mm and $\sim$ 20\% at shorter wavelengths. 
All the spectra are displayed in antenna temperature units.
The main spectroscopic parameters of the transitions observed, and the main
observational parameters, are summarised in Table~\ref{tab_para}. The
spectroscopic parameters have been taken from the Cologne Database for
Molecular Spectroscopy\footnote{https://www.astro.uni-koeln.de/cdms}.

The distances in our sample vary from 120~pc (for the closest object, IRAS 16293) to
$\sim 420$~pc (for the farthest, OMC--2 FIR4). Therefore, the linear scales probed differ
by a factor $\sim 3.5$ at all frequencies, although in reality the difference is less than 
a factor 2 for all objects but OMC--2 FIR4. However, most of our calculations are based on the 
(1--0) transition (see Sect.~\ref{res_col}), whose angular beam size corresponds to a linear 
scale that ranges from a minimum of $\sim 3000$ au for IRAS 16293, to a maximum of 
$\sim 11000$ au for OMC--2 FIR4. According to the description of the sources presented above, 
these scales probe the lower-density envelope in all objects, including OMC--2 FIR4, which
is a protocluster intrinsically larger than the other cores, and thus the isotopic fractions that we
will derive in Sect.~\ref{res_col} are always associated with the envelope of the sources, 
and not with their inner nuclei.

\begin{table}
\centering
\caption{Spectroscopic and observational parameters of the observed lines:
line rest frequency, $\nu$, energy of the upper level, $E_{\rm u}$, line strength,
$S \mu_{\rm ij}^2$, Einstein coefficient of spontaneous emission, $A_{\rm ij}$,
main beam angular size, $\Theta_{\rm MB}$, and main beam efficiency, $B_{\rm eff}$
The spectral parameters have been taken from the Cologne Database for Molecular
Spectroscopy (CDMS; M\"{u}ller et al.~\citeyear{muller01}, 
M\"{u}ller et al.~\citeyear{muller05}).}
\begin{tabular}{rcccccc}
\hline
\hline
Transition & $\nu$ & $E_{\rm u}$ & $S \mu_{\rm ij}^2$ & $A_{\rm ij}$ & $\Theta_{\rm MB}$ & $B_{\rm eff}$ \\
                 & GHz   & K                 & D$^2$                    & s$^{-1}$       &  \arcsec\                  & \\
\hline
${\rm N_2H^+}$ (1--0)  & 93.173 & 4.5 & 104  & $4\cdot 10^{-5}$ & $26$ & 0.80 \\
                        (3--2)  & 279.512 & 27 &  312 & $1\cdot 10^{-3}$    & $9$ & 0.49 \\
${\rm N_2D^+}$ (1--0)  & 77.109  & 4 & 104 & $2\cdot 10^{-5}$ & $32$ & 0.83 \\
                       (2--1)  & 154.217 & 11 & 208 & $2\cdot 10^{-4}$ & $16$ & 0.71 \\
                       (3--2)  & 231.322 & 22 & 312 & $7\cdot 10^{-4}$ & $11$ & 0.59 \\
${\rm ^{15}NNH^+}$ (1--0) & 90.264 & 4 & 35 & $3\cdot 10^{-5}$ & $27$ & 0.80 \\ 
                              (3--2)  & 270.784 & 26 & 104 & $1\cdot 10^{-3}$ & $9$ & 0.51 \\
${\rm N^{15}NH^+}$ (1--0) & 91.206 & 4 & 35 & $3\cdot 10^{-5}$ & $27$ & 0.80 \\ 
                              (3--2)  & 273.609 & 26 & 104 & $1\cdot 10^{-3}$ & $9$ & 0.51 \\                                          
\hline
\end{tabular}
\label{tab_para}
\end{table}
\normalsize

\section{Results}
\label{res}

\subsection{Detected lines and fitting procedure}
\label{res_det}

\begin{table*}
\begin{center}
\caption{Line parameters obtained applying the hyperfine fit procedure (when possible).
In these cases, the ouput parameters are: ${\rm T_{ant}\cdot\tau}$, the peak velocity of the main
hyperfine component, $V_{\rm peak}$, the line width at half maximum, FWHM, the
opacity of the main component, ${\rm \tau_{main}}$, and the integrated line area, $Area$. 
This latter was computed by integrating the spectrum under the channels with
signal above the 3$\sigma$ rms level. The associated uncertainty is computed from the 
propagation of errors: $\sigma \times \Delta v_{\rm res} \times \sqrt{N}$,
where $\sigma$ is the $1 \sigma$ rms is the spectrum, 
$\Delta v_{\rm res}$ is the spectral resolution, and $N$ is the number of channels with signal.
The lines without ${\rm T_{ant}\cdot\tau}$ and ${\rm \tau_{main}}$ could not be fit with
the hyperfine method, so that the associated $V_{\rm peak}$, FWHM, and $Area$ have 
been derived from a Gaussian fit. In these cases, the associated uncertainty on the
$Area$ represents the 1$\sigma$ rms given by the fit procedure. The upper limit on the $Area$ for the 
undetected lines has been calculated assuming a gaussian line with intensity peak equal to 
the 3 $\sigma$ rms level in the spectrum, and FWHM as listed in Col.~5 (for details, see end 
of Sect.~\ref{res_col}). The last column lists the $1 \sigma$ rms level in the spectrum.}
\begin{tabular}{lccccccc}
\hline
\hline
Molecule & Transition & ${\rm T_{ant}\cdot\tau}$ & $V_{\rm peak}$ & FWHM & ${\rm \tau_{main}}$ & $Area$ & rms \\
              &  & K & km s$^{-1}$ & km s$^{-1}$ & - & (K km s$^{-1}$) & (mK) \\
              \hline
              \multicolumn{8}{c}{\textbf{L1544}}\\
\hline
N$_2$H$^+$ & 1--0 & 0.9(0.1) & 6.0(0.1) & 0.3(0.1) & 0.5(0.1) & 5.11(0.04) & 4 \\
N$_2$D$^+$ & 1--0 & 4.9(0.1) & 5.8(0.2) & 0.4(0.6) & 2.7(0.1) & 1.89(0.02) & 3 \\
$^{15}$NNH$^+$ & 1--0 & 0.06(0.02) & 6.9(0.1) & 0.4(0.1) & 1.9(1.8) & 0.023(0.004) & 1  \\
N$^{15}$NH$^+$ & 1--0 & 0.10(0.04) & 6.2(0.1) & 0.3(0.1) & 3.7(2.9) & 0.03(0.01) & 2 \\
\hline
\multicolumn{8}{c}{\textbf{IRAS4 A}}\\
\hline
N$_2$H$^+$ & 1--0 & 0.3(0.1) & 5.9(0.1) & 1.2(0.1) & 0.1(0.01) & 9.10(0.03) & 6 \\
N$_2$D$^+$ & 1--0 & 1.0(0.1) & 7.2(0.1) & 1.2(0.1) & 0.1(0.01) & 1.58(0.02) & 3 \\
           				& 2--1 & 2.2(0.01) & 7.0(0.1) & 1.2(0.1) & 1.2(0.1) & 2.31(0.04) & 6 \\
            				& 3--2 & 1.5(0.1) & 6.0(0.1) & 1.0(0.1)& 1.3(0.2) & 1.17(0.05) & 8 \\
$^{15}$NNH$^+$ & 1--0 & - & 7.3(0.1) & 1.33(0.2) & - & 0.017(0.002) & 1  \\
						& 3--2 &- & 6.8(0.1) & 0.8(0.3)& - & 0.02(0.02) & 7  \\
N$^{15}$NH$^+$ & 1--0 & 0.03(0.01) & 6.2(0.6) & 1.3(2.1) & 0.4(0.1) & 0.043(0.001) & 0.3 \\
           & 3--2 & - & 6.6(0.1) & 0.3(0.1) & - & 0.009(0.002) & 9\\ 
\hline
\multicolumn{8}{c}{\textbf{L1527}}\\
\hline
N$_2$H$^+$ & 1--0 & 0.2(0.1) & 5.6(0.1) & 1.2(0.1) & 0.1(0.01) & 3.47(0.02) & 4 \\
N$_2$D$^+$ $^a$ & 2--1 & - & - & 5.1$^b$ & - & $\leq 0.12$ & 8.3  \\
$^{15}$NNH$^+$ $^a$ & 1--0 & - & - & 3.3$^b$ & - & $\leq 0.04$ & 4  \\
N$^{15}$NH$^+$ $^a$ & 1--0 & - & - & 4.5$^b$ & - & $\leq 0.05$ & 4  \\
\hline
\multicolumn{8}{c}{\textbf{IRAS 16293}}\\
\hline
N$_2$H$^+$ & 1--0 & 0.3(0.1) & 3.0(0.1) & 1.7(0.1) & 0.1(0.01) & 11.78(0.07) & 16 \\
N$_2$D$^+$ & 2--1 & 2.7(0.1) & 4.0(0.1) & 1.0(0.1) & 2.0(0.2) & 2.17(0.01) & 3 \\
           	    & 3--2 & - & 4.5(0.2) & 2.8(0.4)& 0.7(0.1) & 0.72(0.02) & 10 \\
$^{15}$NNH$^+$ $^a$ & 1--0 & - & - & 5.6$^b$ & - & $\leq 0.11$ & 6.4  \\
N$^{15}$NH$^+$ & 3--2 & - & 5.6(0.2) & 2.8(0.5) & - & 0.13(0.02) & 5\\ 
\hline
\multicolumn{8}{c}{\textbf{SVS13-A}}\\
\hline
N$_2$H$^+$ & 1--0 & 0.6(0.1) & 8.2(0.1) & 0.8(0.1) & 0.1(0.003) & 11.40(0.01) & 3 \\
N$_2$D$^+$ & 2--1 & 0.4(0.1) & 8.3(0.1) & 0.6(0.1) & 0.1(0.01) & 0.21(0.02) & 3 \\
          & 3--2 & - & 8.9(0.1) & 0.8(0.1)& - & 0.17(0.02) & 7 \\
$^{15}$NNH$^+$ $^a$ & 1--0 & - & - & 1.9$^b$ & - & $\leq 0.015$ & 2.5  \\
N$^{15}$NH$^+$ & 1--0 & - & 7.6(0.1) & 1.4(0.3) & - & 0.020(0.009)$^{(c)}$ & 1.5 \\
\hline
\end{tabular}\label{tab_res1}
\end{center}
$^{(a)}${Transition undetected: the column density given in Table~\ref{tab_coldens} is an upper limit;}\\
$^{(b)}${Assumed value for FWHM (see Sect.~\ref{res_col}).}\\
\end{table*} 

\begin{table*}
\begin{center}
\caption{Same as Table~\ref{tab_res1} for the "non-evolutionary" sources.}
\begin{tabular}{lccccccc}
\hline
\hline
Molecule & Transition & T$_{ant}\cdot\tau$ & V$_{peak}$ & FWHM & $\tau_{main}$ & Area & rms \\
         &  & K & km s$^{-1}$ & km s$^{-1}$ & - & (K$\cdot$km s$^{-1}$)  & (mK) \\
\hline
\multicolumn{8}{c}{\textbf{L1157--B1}}\\
\hline
N$_2$H$^+$ & 1--0 & 0.004(0.001) & 1.4(0.1) & 3.0(0.1) & 0.1(0.01) & 0.28(0.01) & 2 \\
N$_2$D$^+$ & 1--0 & - & 3.3(0.2) & 1.8(0.5) & - & 0.025(0.014)$^{(c)}$ & 1  \\
$^{15}$NNH$^+$ $^a$ & 1--0 & - & - & 7.5$^b$ & - & $\leq 0.02$ & 1  \\
N$^{15}$NH$^+$ $^a$ & 1--0 & - & - & 9.6$^b$ & - & $\leq 0.03$ & 1  \\
\\
\hline
\multicolumn{8}{c}{\textbf{OMC--2 FIR4}}\\
\hline
N$_2$H$^+$ & 1--0 & 1.4(0.1) & 10.2(0.1) & 1.3(0.1) & 0.1(0.01) & 39.40(0.02) & 6 \\
N$_2$D$^+$ & 2--1 & 0.8(0.1) & 11.0(0.1) & 1.3(0.1) & 0.3(0.6) & 1.24(0.08) & 14 \\
          & 3--2 & - & 10.7(0.02) & 0.9(0.03)& - & 0.80(0.08) & 21 \\
$^{15}$NNH$^+$ & 1--0 & 0.14(0.1) & 11.1(0.1) & 1.2(0.1) & 1.2(0.9) & 0.15(0.01) & 2  \\
N$^{15}$NH$^+$ & 1--0 & 0.14(0.1) & 10.3(0.1) & 1.3(0.1) & 0.8(0.8) & 0.19(0.01) & 1 \\
\hline
\multicolumn{8}{c}{\textbf{CepE}}\\
\hline
N$_2$H$^+$ & 1--0 & 0.2(0.1) & -13.6(0.1) & 1.2(0.1) & 0.1(0.01) & 4.17(0.02) & 4 \\
						   & 3--2 & - & -11.7(0.1) & 4.5(0.1) & - & 4.20(0.02) & 4\\
N$_2$D$^+$  & 1--0 & 0.3(0.1) & -11.1(0.1) & 1.3(0.1) & 1.5(1.0) & 0.29(0.02) & 3 \\
 		 & 2--1 & 1.3(0.1) & -11.2(0.1) & 0.8(0.1) & 2.4(0.3) & 0.79(0.02) & 4 \\
		& 3--2 & - & -12.1(0.3) & 4.6(0.7)& - & 0.28(0.08) & 3 \\
$^{15}$NNH$^+$ & 1--0 & - & -11.9(0.4) & 1.8(1.4) & - & 0.010(0.004) & 2  \\
N$^{15}$NH$^+$ & 1--0 & - & -13.2(0.6) & 4.3(1.2) & - & 0.070(0.001) & 0.3 \\
\hline
%
\end{tabular}\label{tab_res2}
\end{center}
$^{(a)}${Transition undetected: the column density given in Table~\ref{tab_coldens} is an upper limit;}\\
$^{(b)}${Assumed value for FWHM (see text);}\\
$^{(c)}${One or more hyperfine components are under the $3\sigma$ rms level; the $Area$ 
has been computed by summing the integrated intensity of the detected components and the 
expected area of the undetected components (assuming LTE and optically thin conditions).}\\
\end{table*}

We have detected at least one transition of \H\ and \D\ in all the sources of our sample.
In Tables~\ref{tab_res1} and \ref{tab_res2} we list the lines observed and detected in each source.
The \H\ (1--0) line is detected in all targets, while \D\ was observed and detected 
either in the (1--0) and/or in the (2--1) and (3--2) lines. Transitions of \N15\ have 
also been detected in all sources, except L1157--B1 and L1527, while \15N\ lines 
were detected only towards L1544, IRAS 4A, Cep E, and OMC--2 FIR4. 
All the transitions are split into multiple hyperfine components due to the coupling of the 
$^{14}$N nuclear spin with the angular momentum of the molecule. Therefore, we have 
attempted to fit all lines taking the hyperfine structure into account. The method assumes
that the components are Gaussians with the same line width separated in frequency 
according to the laboratory value, and have all the same excitation temperature. 
A detailed description of the method and of the output parameters is given in the CLASS 
manual\footnote{https://www.iram.fr/IRAMFR/GILDAS/doc/pdf/class.pdf}.
For the \H\ and \D\ lines, the method has given generally good results, while it was not 
appropriate for several \15N\ and \N15\ lines, mainly because of the faintness of the secondary 
components. In particular, for two lines, \N15\ (1--0) in SVS13-A and \D\ (1--0) in L1157--B1,
the main component was clearly detected, but one or more satellites were undetected. Therefore,
we have fitted the main component with a single Gaussian, and then calculated 
the expected contribution of the undetected hyperfine components assuming LTE 
and optically thin conditions. The final integrated area for these lines is the sum of these
contributions (see Tables~\ref{tab_res1} and \ref{tab_res2}). 
In Cols. 3--8 of Tables~\ref{tab_res1} and \ref{tab_res2} 
we show the line parameters derived with the methods mentioned above.
The Tables also give the $1\sigma$ rms level in the spectrum (Col.~9), used for non-detected
lines to compute the upper limits on the column densities. Transitions not shown in the Tables 
were not observed. The spectra of all lines detected are shown in the on-line Appendix-A.

\subsection{Derivation of the total column densities}
\label{res_col}

From the line parameters, we have calculated the total column densities $N$ of \H , \D ,
\15N\ and \N15 . 
Because different sources have been detected in different lines, and, above
all, not all targets have been observed in the same lines, to be consistent, 
we have derived $N$ from the (1--0) line for all species, when possible.
This approach was suggested by the fact that the (1--0) line is detected in
almost all species and targets, and because the isotopic ratio derived by dividing
column densities obtained from the same transition are independent from the 
excitation temperature. The method has been used successfully
for L1544, IRAS 4A, and L1157--B1.
In the other targets, this method can not be applied because the (1--0) line
was either not observed or not detected in all molecules. For
example, in IRAS 16293, the (1--0) transition is detected in \H\ and \15N but neither 
in \D\ nor in \N15. For these two molecules, we have detected the (2--1) and (3--2) 
lines (see Table~\ref{tab_res1}). In this case, the column densities were derived 
adopting the transition with the best signal-to-noise ratio in each molecule, assuming
a wide (but reasonable) range of excitation temperatures of 5 -- 30~K. For L1544, 
we have assumed a more realistic temperature range of 5 -- 15~K. 

The fit to the hyperfine structure gives an estimate of the optical depth of the main component, 
$\tau_{main}$. For lines with $\tau_{main}\leq 0.5$, we have computed $N$ from the total
line integrated area according to Eq.~(A4) in Caselli et al.~(\citeyear{caselli2002b}), valid for 
optically thin lines. The integrated areas have been computed from 
the spectra in antenna temperature units, and then converted to main beam brightness 
temperature units using the main beam efficiencies reported on the IRAM--30m 
website\footnote{http://www.iram.es/IRAMES/mainWiki/Iram30mEfficiencies}.
The optically thin approach is good for almost all the detections 
(see Tables~\ref{tab_res1} and \ref{tab_res2}). Some \D\ lines have $\tau_{main}> 0.5$, 
but poorly constrained ($\Delta \tau_{main}/\tau_{main}\geq 0.3$), 
so that we have derived the column densities using the optically thin approach as well.
For the few optically thick lines and with opacity well-constrained, the column density 
has been calculated by multiplying the value derived in optically thin approximation
by the factor $\tau/(1-e^{-\tau})$ (where $\tau$ is the total opacity of the line). 
Finally, for the few \15N\ transitions in which the secondary components are not
detected, we have used only the integrated intensity of the main one, assuming 
optically thin conditions, hence in these cases (see Tables~\ref{tab_res1} and \ref{tab_res2}) 
the derived $N$ has to be regarded as lower limits.

The column densities should be corrected for the filling factor to compare them
in a consistent way, but in most cases we do not know the emission size of the transitions 
observed, therefore we do not have a direct estimate of the filling factor. However, as shown 
in Castets et al.~(\citeyear{castets}), in IRAS 16293 the source size in \H\ (1--0) can be as 
extended as $\sim 30$\asec, i.e. bigger than the beam size of all transitions observed. 
This is reasonably the extension of the emission also in the other sources, because it
is well-known that the \H\ (1--0) line traces the core envelope. 
For the (1--0) lines of the isotopologues, we have assumed the same emission size as for
\H , and in the few cases in which only the (2--1) or (3--2) line was detected, we have 
assumed that the emission fills the telescope beam. The different 
transitions do not have the same critical density, which is $\sim 10^{5}$ \cmc\ for the (1--0) line, 
and $\sim 10^{6}$ \cmc\ for the (3--2) line (see Lique et al.~\citeyear{lique} for the collisional
coefficients). Hence the angular size of the (3--2) line emitting
region is likely smaller than that of the (1--0) line. However, because the telescope
beam size of the (3--2) transition is also smaller (Table~\ref{tab_para}) than that of the (1--0)
line, in the absence of high-angular resolution observations it is reasonable to assume 
that even the higher excitation transitions fill their (smaller) telescope beam sizes. Moreover,
the isotopic ratio is derived comparing the (1--0) and (3--2) transitions only for IRAS 16293 
and CepE (this latter does not even belong to the "evolutionary" sample). 
Hence, the two isotopic ratios obtained from different excitation transitions, although 
suffering from higher uncertainties, are not crucial for the trends and the overall conclusions 
of the work.

The total molecular column densities are given in Table~\ref{tab_coldens}, and the associated 
isotopic ratios are listed in Table~\ref{tab_ratios}. For each source, the estimates have been 
made assuming a reasonable range of excitation temperatures (5--30~K for all sources but 
L1544, for which a more realistic range 5 --15~K is used). We stress that the temperature 
assumed is irrelevant when the \r1415\ is derived from the (1--0) line of both isotopologues,
but even when derived from different excitation transitions, the difference is within the
uncertainties (typical uncertainty in between $20\%$ and $50\%$, see Table~\ref{tab_coldens}). 
The \H\ average column densities are of the order of $N({\rm N_2H^+})\sim10^{14}$\cmq , 
while $N$(\D ) range from $\sim 10^{11}$ up to $\sim 10^{13}$\cmq , and both 
$N$(\15N ) and $N$(\N15 ) are in the range $10^{11} - 10^{12}$\cmq . 

Finally, for the sources undetected in \D, \15N\ and \N15, we have calculated the upper
limits on the column density from the upper limit on the line integrated area. This
was evaluated as the area of a single Gaussian having peak equal to the $3 \sigma$ rms
level of the spectrum. To compute the integrated area, however, one needs an estimate 
of the line width. This latter was evaluated as the width obtained from a Gaussian fit
to the other, well-detected lines. In those cases in which the fit with a single Gaussian does
not give reasonable results, we have computed the line width by applying to the line
width obtained from the hyperfine structure a factor derived from the lines with good
fit results with both methods.  

\section{Discussion}
\label{discu}

\begin{figure}
\begin{center}
\caption{D/H ratio derived from the column density ratio \H /\D\ (see Sect.~\ref{res_col})
against the \r1415\ ratio derived from both \15N\ (red and green symbols) and \N15\ 
(blue and yellow symbols).
}
{\includegraphics[angle=0, width=8.5cm]{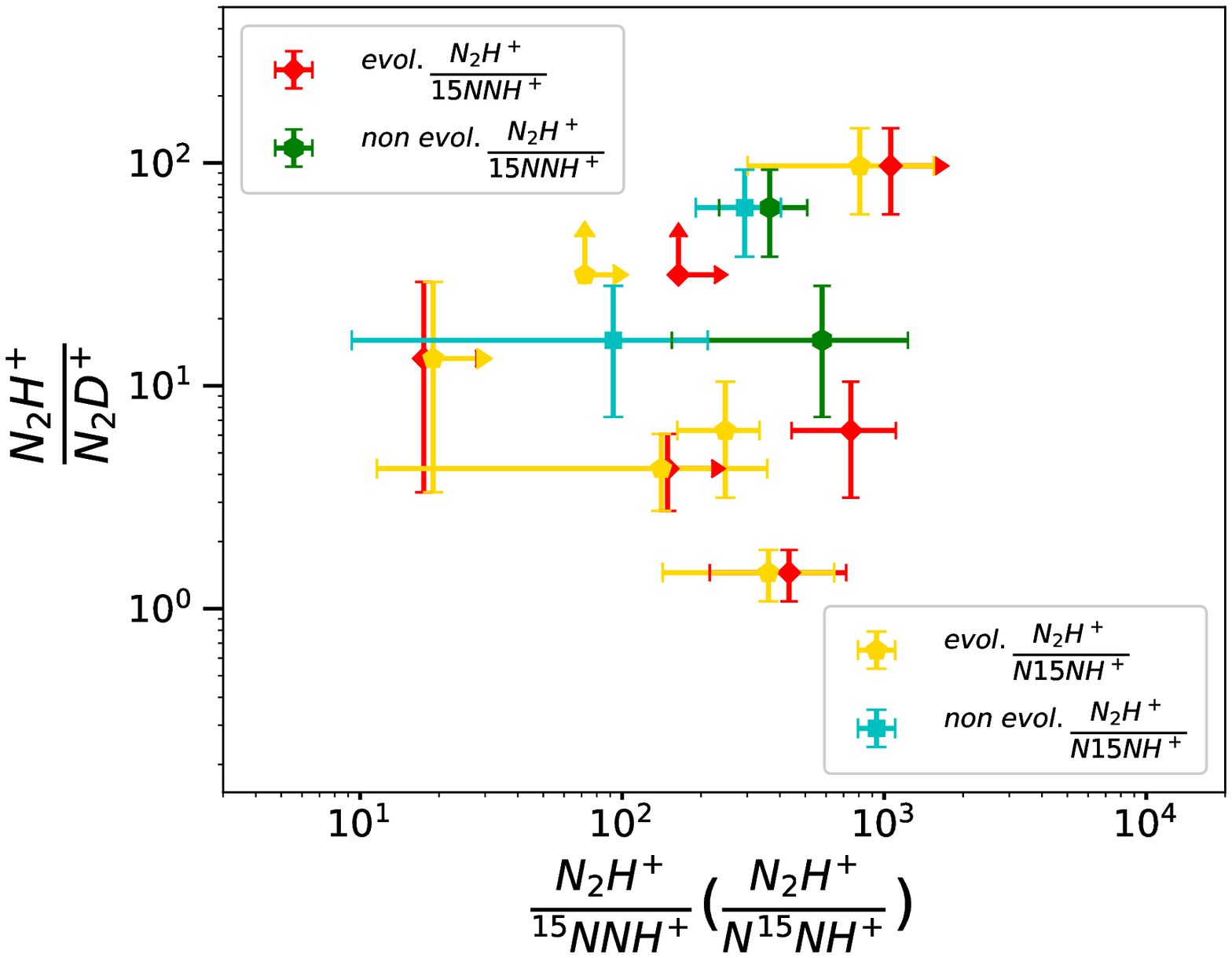}}
\label{fig_Dvs15N}
\end{center}
\end{figure}

In Fig.~\ref{fig_res} we show the trends of the three isotopic ratios. We indicate in
different colours the low-mass sources with well established evolutionary stage, namely 
L1544, IRAS 4A, IRAS 16293, L1527, and SVS13-A, and the other objects. We have 
ordered the "evolutionary sample" with increasing time. Although L1157--B1 does 
not belong to the evolutionary sample of cores, as explained in Sect.~\ref{obs} the 
measurement of the fractionation in this object reflects that of the diffuse gas of the
cloud before the condensation into a dense pre--stellar core, hence we have placed it
in the evolutionary sequence before L1544. The two hot-corinos IRAS 4A and IRAS 16293 
are placed closer than the other objects because there is no evidence that can
indicate a different age of these two objects. 
Moreover, in Fig.~\ref{fig_Dvs15N} we plot the \H /\D\ versus \H /\15N\ and \H /\N15\ 
ratios, in order to highlight a possible correlation between the D and 15N fractionation.

The most direct results emerging from Figs.~\ref{fig_res} and \ref{fig_Dvs15N} are: 
\begin{itemize}
\item [(1)] the \H /\D\ decreases, as expected, from the diffuse cloud stage
(represented by L1157--B1) to the pre--stellar core stage, represented by L1544, by
a factor $\sim 10$. Then, it increases monotonically in the more evolved stages, again
by a factor $\sim 10$ in the Class 0 objects, and by a factor $\sim 100$ in the more
evolved Class I object SVS13-A;
\item [(2)] the \r1415\ is quite uniform across the sample, which indicates that time is globally 
irrelevant in the fractionation of nitrogen in \H. 
Moreover, in the four objects in which both \15N\ and \N15\ are detected, namely 
L1544, IRAS 4A, CepE, and OMC--2 FIR4, the \r1415\ ratio is the same, within the uncertainties,
when derived from the two isomers.
The only exception is IRAS 4A, in which they are different by a factor 2 (see Table~\ref{tab_ratios}
and Fig.~\ref{fig_res}). This indicates that the two isotopologues tend to follow the same 
chemical pathway, although the peculiar case of IRAS 4A deserves to be investigated
further, possibly with the help of higher sensitivity and higher spectral resolution observations.
Moreover, one can note that the \r1415\ ratio derived from \15N\ is systematically higher than 
that computed from \N15\, which suggests that the \N15\ is more abundant than \15N. This 
difference could be due to the slightly different zero-point energy of the ground state level of 
the two molecules, which would favour the formation of \N15\ (see e.g.~Terzieva \& 
Herbst~\citeyear{teh}). However, all this needs to be supported and confirmed by a
higher statistics;
\item [(3)] the two isotopic ratios D/H and \r1415\ are not related, as it can be seen from 
Fig.~\ref{fig_Dvs15N}. This result is a direct consequence of the fact that the \r1415\ 
is independent of evolution, opposite to the D/H ratio. Therefore, we do not confirm the possible
anti-correlation claimed by Fontani et al.~(\citeyear{fontani2015}) in massive star-forming
cores, although even in that study the anti-correlation was tentative due to the large
dispersion of the points. Moreover, our dataset and that of Fontani et al.~(\citeyear{fontani2015})
are different in many aspects:
the linear size of the observed region is much more extended in the high-mass objects
of Fontani et al.~\citeyear{fontani2015} (located at more than 1~kpc), and the evolutionary 
timescales in massive star-forming cores are significantly shorter;
\item [(4)] inspection of Fig.~\ref{fig_res} also shows that the D/H ratio in CepE is consistent 
with that measured in IRAS 16293 and IRAS 4A, which suggests that low- and 
intermediate-mass class 0 protostars have a comparable deuteration in \H . Of course,
this finding needs to be supported by a more robust statistics.
 \end{itemize}

That the core evolution is basically irrelevant for the \r1415\ ratio was concluded also by
Fontani et al.~(\citeyear{fontani2015}) and Colzi et al.~(\citeyear{colzi}) from observations
of several molecular species (\H, CN, HNC, and HNC) towards high-mass star-forming cores belonging
to different evolutionary stages. Both our study and those performed in high-mass star-forming
regions indicate that the enrichment of $^{15}$N is unlikely to happen at core scales during 
the formation of stars of all masses. Therefore, we speculate that the $^{15}$N abundance 
enhancement measured in pristine Solar system material should be due to chemical processes 
occurred locally, perhaps at the scale of the protoplanetary disk, and not in the extended surrounding 
envelope. The same conclusion was provided also by Kahane et al.~(submitted to ApJ) 
towards OMC--2 FIR4. We will discuss this point further in Sect.~\ref{individual}.
Interestingly, cometary-like \r1415\ have indeed been measured by Guzm\'an et al.~(\citeyear{guzman})
in a sample of protoplanetary disk, suggesting a chemical link. However, the study of Guzm\'an et 
al.~(\citeyear{guzman}) is focused on HCN and HNC isotopologues, hence obviously what happens 
in \H\ and its isotopologues could be completely different and still needs to be investigated in disks.
Interestingly, our \r1415\ ratios are consistent, within the uncertainties, with those measured
from ammonia and NH$_2$D in a sample of dense and young star-forming cores (e.g.~Gerin et 
al.~\citeyear{gerin2009}, Daniel et al.~\citeyear{daniel2013}). This would suggest that the
\r1415\ ratio in the envelope of star-forming cores does not depend even on the molecule used. 
However, both our work and those on ammonia mentioned above are based on a small
statistics, thus only observations on larger, carefully selected samples of star-forming 
cores, like that studied in this work, will confirm this possibility.

\subsection{Comments on individual sources}
\label{individual}

\noindent

\begin{figure}
\begin{center}
\caption{Isotopic ratios H/D (top panel) and \r1415\ (bottom panel) 
obtained from the \H\ isotopologues from the method described in Sect.~\ref{res}.
The points represent the values obtained for an average excitation temperature of 17~K, and the
uncertainty is derived from the scatter calculated in the "reasonable" temperature range 
of 5 -- 30~K (see Table~\ref{tab_coldens} and Sect.~\ref{res_col} for details).
In the top panel, red diamonds represent the sources of the "evolutionary" sample (see Sect.~\ref{intro}), 
while the other objects are represented by green hexagons. L1157--B1 has been
placed before L1544 on the x-axis because the fractionation obtained in this object
is associated with the diffuse gas before the passage of the shock (Codella et al.~\citeyear{codella2013}). 
The points with an upward arrow represent the lower limit of the ratios derived 
from the upper limits of the column density. 
In the bottom panel, red and yellow diamonds indicate the isotopic fraction obtained
from \N15\ and \15N, respectively, for the sources of the "evolutionary sample", while for
the others (CepE and OMC--2 FIR4) we use blue and green symbols, respectively.}
{\includegraphics[angle=0, width=9cm]{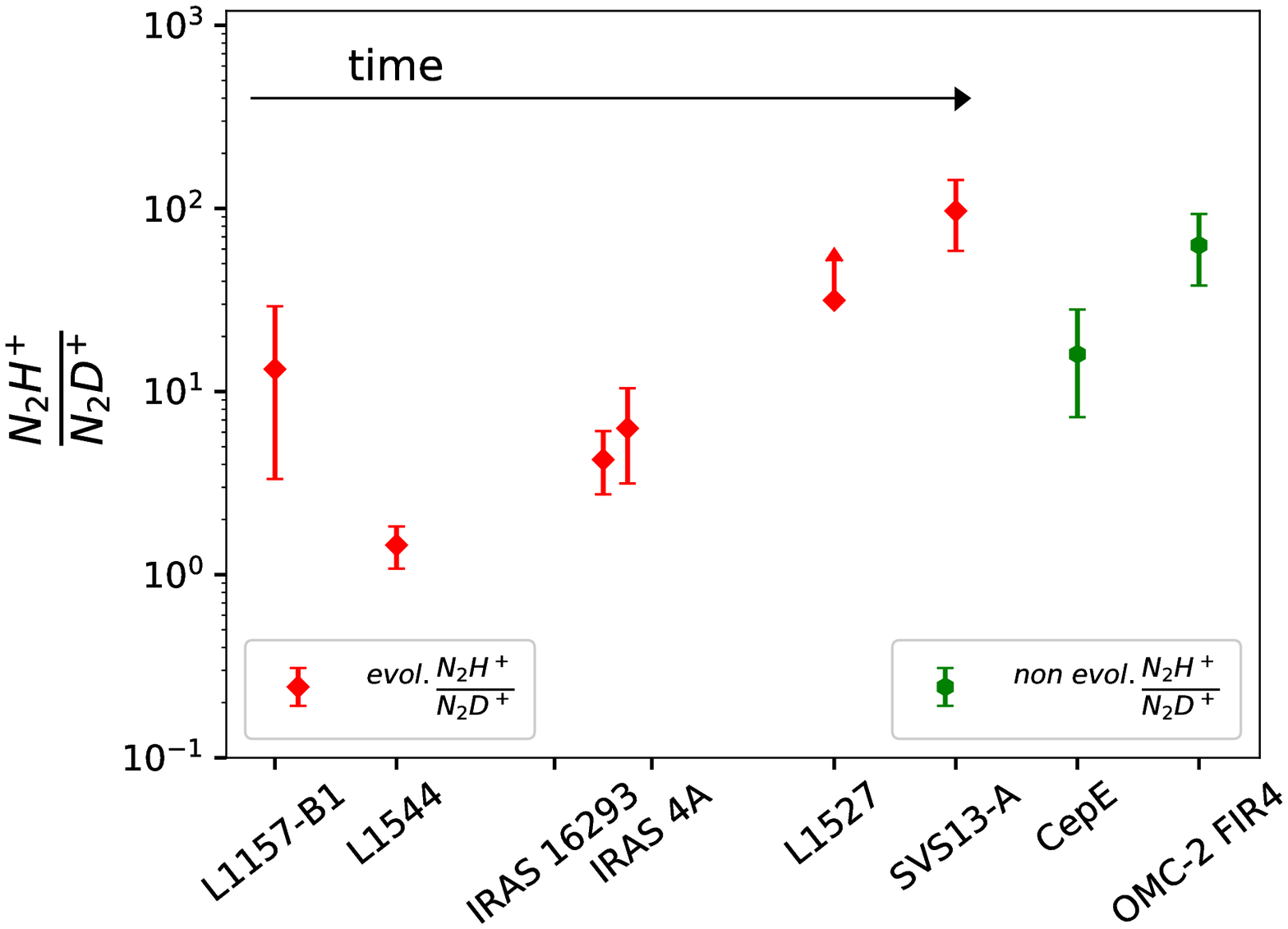}}
{\includegraphics[angle=0, width=9cm]{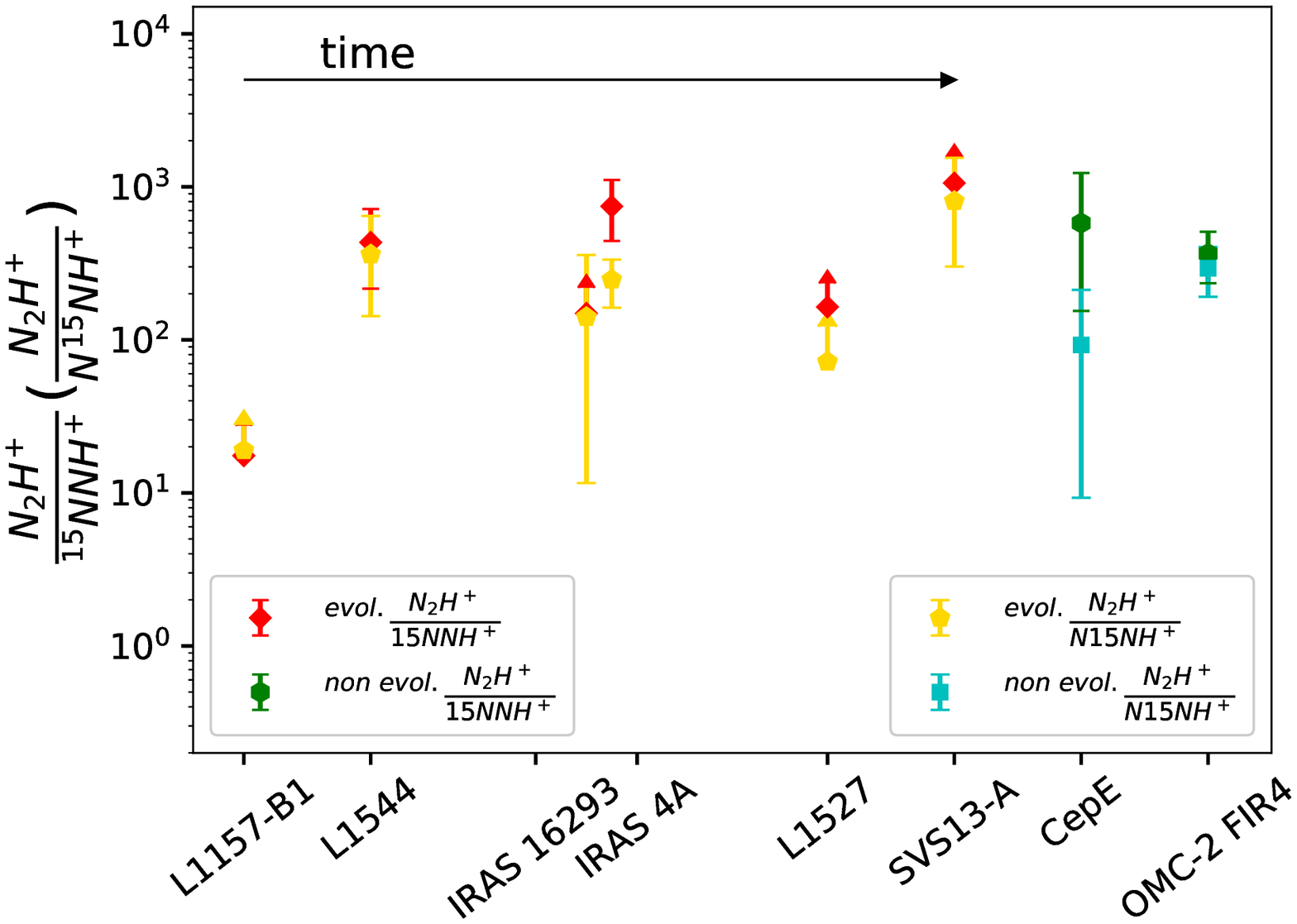}}
\label{fig_res}
\end{center}
\end{figure}

{\bf L1544:} the \r1415\ ratio in \H\ was already derived by Bizzocchi et al.~(\citeyear{bizzocchi})
through a non-LTE method using a Bonnor-Ebert sphere model for the source. This method
provided \r1415\ $=1000\pm 200$, while with our LTE method we derive \r1415\ in the range 275 -- 408
and 228 -- 341 from $^{15}$NNH$^+$ and N$^{15}$NH$^+$. However, we stress that within
the uncertainties, the values are still marginally consistent, as we can note from 
Fig.~\ref{fig_res}. Also, even with the value given by Bizzocchi et al.~(\citeyear{bizzocchi}) the 
global evolutionary trend on the \r1415\ in Fig.~\ref{fig_res} would not change significantly.
On the other hand, the deuterated fraction of $\sim 0.24$ previously calculated by Caselli et 
al.~(\citeyear{caselli2002b}) with the same approach adopted by us, is perfectly consistent 
with our estimates ($\sim 0.25 - 0.33$).

\noindent

{\bf SVS13-A:} we find different values for the \r1415\ derived from \15N\ and \N15 . 
In particular, the lower limit derived from \15N\ is larger ($\sim 700-1400$, see Table~\ref{tab_ratios}) 
than the upper limit calculated from \N15\ ($\sim 360-700$), as well as any other \r1415\ measured 
in our survey. Please note that in Fig.~\ref{fig_res}, we have shown the intermediate values
(1050 and 530, repectively).
Because SVS13-A is the only class I object of the sample, if confirmed, this result would point to
a different chemical behaviour between the two $^{15}$N isotopologues with time, which is not
predicted by the most recent theoretical models (Roueff et al.~\citeyear{roueff}). Hence, it will be
worth investigating this result with higher sensitivity observations, and possibly a larger statistics
of similar objects.
%

\noindent

{\bf OMC--2 FIR4:} as stated in Sect.~\ref{obs}, this protocluster is the closest analogue of
the environment in which our Sun is thought to have been born. The nitrogen fractionation
has been extensively studied by Kahane et al.~(submitted to ApJ), who found a good
agreement between the present-day local interstellar \r1415 , and the \r1415\ measured
from several molecules (HCN, HNC, HC$_3$N, CN). Our results (Table~\ref{tab_ratios})
are perfectly consistent with this finding, indicating a remarkable uniformity of this ratio
independently from the molecule used, and a further indication that the large-scale gas
is not enriched in $^{15}$N, as concluded by Kahane et al.~(\citeyear{kahane}).

\begin{table*}
\begin{center}
\caption{Total column densities of \H , \D , \15N , and \N15\ calculated as explained in Sect.~\ref{res}.
For each species, we have assumed a range in excitation temperature, \Tex , based on the
reference papers in the footnotes. In Col.~2, we indicate the transition from which the total 
column density of each species has been derived.}
\label{tab_coldens}
\begin{tabular}{lcccccccc}
\hline
\hline
Source & Transition & \Tex\ & $N$(N$_2$H$^+$) & $N$(N$_2$D$^+$) & $N$($^{15}$NNH$^+$) & $N$(N$^{15}$NH$^+$) \\
           &         & K & 10$^{13}$cm$^{-2}$ & 10$^{13}$cm$^{-2}$ & 10$^{10}$cm$^{-2}$ & 10$^{10}$cm$^{-2}$\\
\hline
L1544 & 1--0 & 5 - 15 & 32.6 - 12.9 & 18.5 - 11.6 &  62.7 - 37.2 & 75.3 - 44.4  \\
IRAS 4A & 1--0 & 5 - 30 & 53.9 - 26.3 & 6.3 - 6.5  & 54.7 - 52.0 & 165 - 156 \\
IRAS 16293 & 1--0 & 5 - 30 & 7.0 - 3.4 & - & $\leq$ 35.4 - 33.7 & - \\
		 & 2--1 & 5 - 30  & -  & 2.0 - 0.7 & - & - \\
		 & 3--2 & 5 - 30 & - & - & - & 393 - 12.9   \\
L1527 & 1--0 & 5 - 30 & 20.6 - 10.0 & - & $\leq$ 128.6 - 122.4 &  $\leq$ 142.7 - 135.1\\
	 & 2--1 & 5 - 30 & - & $\leq$ 0.48- 0.49 & - & - \\
SVS13-A & 1--0 & 5 - 30 & 6.8 - 3.3 & - & $\leq$ 4.8 - 4.6 &  6.3 - 6.0 \\ 
		 & 2--1 & 5 - 30 & 	- & 0.08 - 0.03 & - & - \\
L1157--B1 & 1--0 & 5 - 30 & 1.7 - 0.8 & 0.1 - 0.1 & $\leq$ 61.1 - 58.2 & $\leq$ 80.6 - 76.3 \\
CepE		& 1--0 & 5 - 30 & 2.5 - 1.2 & 0.11 - 0.12 & 3.2 - 3.1 & - \\
		& 3--2 & 5 - 30 &  - & - & - & 212 - 6.9 \\
OMC--2 FIR4 & 1--0 & 5 - 30  & 19.8 - 9.6 & - &  40.9 - 38.9 & 51.1 - 48.3 \\
		    & 2--1 & 5 - 30 & - & 0.4 - 0.1 & - & - \\
\hline
\end{tabular}
\end{center}
\end{table*}

\begin{table}
\begin{center}
\label{tab_ratios}
\caption{D/H and \r1415\ isotopic ratios calculated as explained in Sect.~\ref{res_col}.}
\begin{tabular}{lcccc}
Source  & \Tex\ & ${\rm \frac{N_2H^+}{N_2D^+}}$ & ${\rm \frac{N_2H^+}{^{15}NNH^+}}$ & ${\rm \frac{N_2H^+}{N^{15}NH^+}}$  \\
 &  K &  &  &  \\
\hline
L1157--B1 & 5 - 30 & 18.4 - 8.1 & 27.2 - 13.9$^l$  & 20.6 - 10.6$^l$ \\
L1544 & 5 - 15 & 1.8 - 1.1  & 520 - 347 & 433 - 290 \\
IRAS 4A & 5 - 30 & 8.6 - 4.1 & 986 - 505 & 327 - 168  \\
IRAS 16293 & 5 - 30 & 3.5 - 4.8 & 197 - 101$^l$ & 18 - 264  \\
L1527 &  5 - 30 & 42.9 - 20.4$^l$ & 160 - 168$^l$ & 70 - 74$^l$  \\
SVS13-A & 5 - 30 & 84 - 110 & 1400 - 717$^l$ & 1065 - 548 \\
CepE &  5 - 30 & 22 - 10 & 767 - 392 & 12 - 173  \\
OMC--2 FIR4 &  5 - 30 &  52 - 74 & 484 - 247  & 388 - 199 \\
\hline
\end{tabular}
\end{center}
$^l$ lower limit; \\
\end{table}


\section{Conclusions}
\label{conc}

In the framework of the IRAM-30m Large Program ASAI (Lefloch et al.~in prep.), we
have analysed the rotational transitions of \H , \D , \15N , and \N15\ in order to
investigate if (and how) the isotopic fractions H/D and \r1415\ vary during the
formation of a Sun-like star. We have confirmed in a robust way that the \H /\D\ is
a clear evolutionary indicator in low-mass star formation, because this ratio reaches the
minimum value in L1544 (around $\sim 3 - 5$), i.e. during the pre--stellar core phase at
the onset of the gravitational collapse, and then increases with time monotonically by 
two orders of magnitude during the protostellar phase, as expected from chemical models. 
On the other hand, our data do not indicate an evolutionary trend
for the \r1415\ ratio, and show clearly that the two isotopic ratios H/D and \r1415\ are basically
independent. Therefore, our study confirms previous claims obtained in high-mass star-forming
cores that the two ratios are not influenced by the same physical/chemical processes. 
Also, because our findings demonstrate that the
$^{15}$N enrichment is a process unlikely to happen in the envelope of forming stars of
all masses, the enrichment measured in pristine Solar system material should be caused
by chemical processes occurred locally at the scale of the protoplanetary disk.

\vspace{0.2cm}
\noindent
{\it Acknowledgments.}  We are grateful to the anonymous referee for the useful and 
constructive comments that have improved the paper.
We thank the IRAM staff for the precious help during the observations. 
F.F. and C.Codella akcnowledge financial support from the Italian Ministero dell'Istruzione, 
Universit\`a e Ricerca through the grant Progetti Premiali 2012 - iALMA (CUP C52I13000140001).
C.Ceccarelli acknowledges the funding from the European Research Council (ERC) under the 
European Union's Horizon 2020 research and innovation programme, project DOC contract 741002.

\let\oldbibliography\thebibliography
\renewcommand{\thebibliography}[1]{\oldbibliography{#1}
\setlength{\itemsep}{-1pt}} 

{}

\normalsize

\clearpage

\normalsize

\normalsize

\end{document}